\documentclass[twocolumn,showpacs,preprintnumbers,amsmath,amssymb,prb]{revtex4}

\usepackage{graphicx}
\usepackage{dcolumn}
\usepackage{bm}

\newcommand{\ttise}{1\textit{T}-TiSe$_2$}

\begin{document}

\preprint{APS/123-QED}
\bibliographystyle{prsty}

\title{Temperature dependent photoemission on 1$T$-TiSe$_{2}$: Interpretation within the exciton condensate phase model}

\author{C. Monney}
\email{claude.monney@unifr.ch}
\author{E.F. Schwier}
\author{C. Battaglia}
\author{M.G. Garnier}
\author{N. Mariotti}
\author{C. Didiot}
\author{H. Beck}
\author{P. Aebi}

\affiliation{%
Institut de Physique, Universit\'e de Fribourg, CH-1700 Fribourg, Switzerland
}%

\author{H. Cercellier}
\author{J. Marcus}

\affiliation{%
Institut N\'eel, CNRS-UJF, BP 166, 38042 Grenoble, France
}%

\author{H. Berger}

\affiliation{%
Institut de Physique de la Mati\`ere Condens\'ee, EPFL, CH-1015 Lausanne, Switzerland
}%

\author{A.N. Titov}

\affiliation{%
Institute of Metal Physics UrD RAS, Ekaterinburg, 620219, Russia\\
Institute of Metallurgy UrD RAS, Amundsen St. 101, Ekaterinburg, 620016, Russia
}%

\date{\today}

\begin{abstract}
The charge density wave phase transition of 1$T$-TiSe$_{2}$ is studied by angle-resolved photoemission over a wide temperature range. An important chemical potential shift which strongly evolves with temperature is evidenced. In the framework of the exciton condensate phase, the detailed temperature dependence of the associated order parameter is extracted. Having a mean-field-like behaviour at low temperature, it exhibits a non-zero value above the transition, interpreted as the signature of strong excitonic fluctuations, reminiscent of the pseudo-gap phase of high temperature superconductors. Integrated intensity around the Fermi level is found to display a trend similar to the measured resistivity and is discussed within the model.
\end{abstract}

\pacs{79.60.Bm,71.27.+a,71.35.Lk,71.45.Lr}
\maketitle

\section{Introduction}

The transition metal dichalcogenides (TMDC) belong to a class of quasi two-dimensional compounds famous for their charge density wave (CDW) phases \cite{FlorianThesis,MotizukiBook,AebiJElectSpect}. Due to their layered structure, they can be easily intercalated by foreign atoms in their so-called Van der Waals gap, providing a chemical parameter for tuning new phenomena. In this way, for instance superconductivity can be enhanced or suppressed, sometimes in competition with CDW phases \cite{MorosanNature,FangTaS2}.

\begin{figure}
\centering
\includegraphics[width=8.5cm]{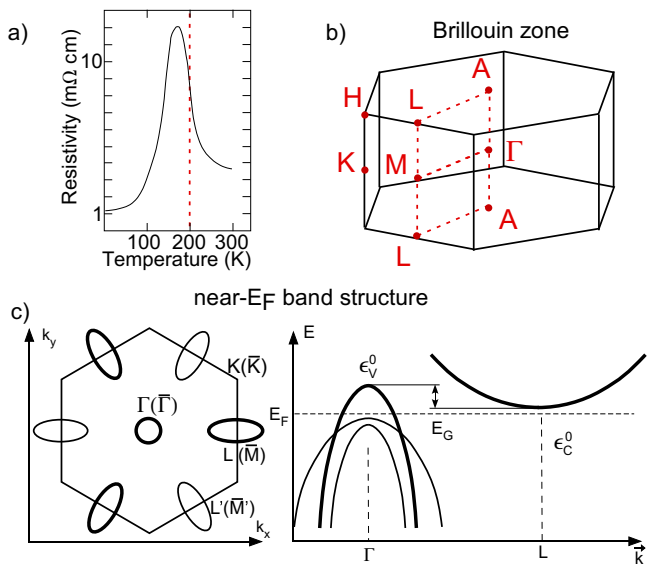}
\caption{\label{PRB_Tscan_fig_1} 
(a) In-plane ($\perp c$) resistivity of \ttise\ as a function of temperature\cite{LevyResist}. (b) Brillouin zone of \ttise\ with its high symmetry points. (c) Left: constant energy map near $E_F$ of \ttise\ in planes parallel to the $\Gamma MK$ plane, with electron pockets at $L(\bar{M})$ produced by the conduction band. In parenthesis the surface Brillouin zone notation is indicated for high symmetry points. Right: the valence bands have their maxima at $\Gamma$ and the conduction band has its minimum at $L$.
}
\end{figure}

Among the TMDC, \ttise\ turns out to be an interesting and enigmatic material. At the critical temperature of $T_c\simeq 200$K, the system undergoes a second-order phase transition, characterized notably by a peaking resistivity\cite{DiSalvoSuperLatt,LevyResist} (see Fig. \ref{PRB_Tscan_fig_1} (a))  and a phonon softening\cite{HoltPhonSoft} at $T_c$. The electronic band structure and its elementary excitations can be determined by angle-resolved photoemission spectroscopy (ARPES). At room-temperature (RT), ARPES evidenced two main contributions near the Fermi energy $E_F$, namely a valence band (of Se$4p$ character) and conduction bands (of Ti$3d$ character), whose relative energy positions are still controversial \cite{RaschAdsorption,CercellierPRL}. At low temperature (LT), intense backfolded bands, characteristic of the CDW, appear. The origin of the CDW is not completely settled yet and resists to conventional explanations. Indeed,  the \ttise\ Fermi surface (FS) topology does not favor nesting, since no large parallel portions of FS are present \cite{ZungerDFT}. A band Jahn-Teller effect \cite{HuguesBJT} has been proposed as an alternative mechanism, relying on the fact that at the transition a periodic lattice distortion develops, which results in a tendency of the system to pass from the 1\textit{T} (octhedral environment of the transition atom) to the 2\textit{H} (prismatic environment) polytype. A third explanation, that is developed hereafter, is the exciton condensate phase.

This phase, originally denominated as the \textit{excitonic insulator phase}, appeared in the mid-1960s as a theoretical prediction \cite{KeldyshEI,JeromeBasis}. In its simplest version, its basic ingredients are a single valence and a single conduction band, having a semimetallic or semiconductor configuration. Then, if the overlap or the gap between them is small enough, bound states of holes and electrons, called excitons, will condense in a macroscopic state and drive the system into a new ground state, provided the temperature is sufficiently low. The CDW arises naturally from the coupling between the valence and conduction bands, opening a gap between them at LT and transforming the semimetallic or semiconducting configuration into an insulating one. In other words, the presence of condensed excitons in the system creates the CDW as a purely electronic process. 

In the case of \ttise, the situation is more complicated. Three symmetry equivalent conduction bands having their minima at the border of the Brillouin zone (BZ) (at the $L$ point, see Fig. \ref{PRB_Tscan_fig_1} (b)) are coupled to the valence band having its maximum at the center of the BZ (the $\Gamma$ point). The main difference with the basic excitonic insulator phase is that the electron-hole coupling does not shift all the conduction bands, providing states which are unperturbed by the transition and therefore tempering the insulating character of the transition \cite{MonneyPRBTheo}.

Among the recent ARPES studies on \ttise, Pillo {\it et al.} inferred the existence of a small indirect gap and a conduction band in the unoccupied states and supported the exciton condensate phase scenario \cite{PilloTiSe2}. Kidd {\it et al.} also evidenced a small indirect gap with a conduction band in the unoccupied states. They relied on a combination of an electron-hole coupling and a Jahn-Teller effect as the origin of the CDW phase \cite{KiddTiSe2}. Rossnagel {\it et al.} deduced remarkable shifts of the valence and conduction bands in the temperature range of $100$K$\leq T\leq 300$K, but without being able to determine the nature of the gap \cite{RossnagelTiSe2}. On this basis, they also gave a simple qualitative explanation of the peak in the resistivity of \ttise. In their conclusions, they rather endorse the Jahn-Teller effect. 
In our recent publications, we defend the thesis of the excitonic insulator phase as the origin of the CDW phase of \ttise\ and we infer a small indirect overlap of the valence and conduction bands in the normal phase, i.e. in the absence of excitonic effects \cite{CercellierPRL,MonneyPRBTheo}.

In this paper, we present high resolution ARPES measurements of \ttise\ aimed at extracting a detailed temperature dependence over a wide temperature range. It evidences strong shifts of the backfolded valence band and of the conduction band, which we are able to relate to the combined effects of the order parameter characterizing the exciton condensate phase and of a chemical potential shift. Moreover, in photoemission spectra, relevant intensity features following closely the shape of the temperature dependent resistivity are found. 
Finally, at the lowest temperature achieved here, we identify for the first time a new structure in the conduction band which we discuss in relation to the electron-phonon coupling. Section II presents the experimental data whereas Sections III and IV give a detailed discussion before finishing with Section V. 
\\

\section{Experimental data}

The photoemission intensity maps presented here were recorded using linear $p$-polarized HeI$\alpha$ radiation at 21.2 eV and using an upgraded Scienta SES-200 spectrometer with an overall energy resolution better than 10 meV. A liquid helium cooled manipulator having an angular resolution of $~0.1^\circ$ was used, with a temperature stability $<5$K. \ttise\ samples were cleaved \textit{in-situ}, in a base pressure in the low $10^{-11}$ mbar, ensuring a high longevity of the sample surface. Photoemission spectra were recorded from 13K to RT. At the end of the measurements, the sample was cooled again to 13K and comparable spectra were recorded again, confirming the stability of the surface. Reference spectra of polycrystalline gold evaporated on the same sampleholder as \ttise\ were recorded for determining the Fermi level position. At the excitation energy of 21.2 eV, at the border of the BZ, initial states close to the $L$ point are probed (see the BZ depicted in Fig. \ref{PRB_Tscan_fig_1}(b) for situating high symmetry points). In the following, we will use the surface BZ notation, $\bar{M}$, for such measurements. 

\begin{figure}
\centering
\includegraphics[width=9cm]{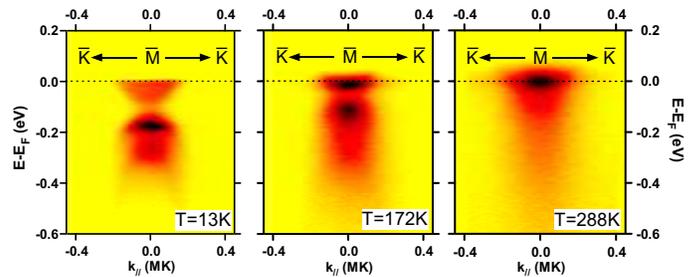}
\caption{\label{PRB_Tscan_fig_2} 
(color online) Photoemission intensity maps, as false colour plots (dark colours represent strong intensity), measured along $\bar{\Gamma}\bar{K}$ (see text for an explanation of this notation) at three different temperatures corresponding, from left to right, to situations well below $T_c$, around $T_c$ and well above $T_c$ respectively.
}
\end{figure}

Fig. \ref{PRB_Tscan_fig_1}(c) presents the schematic near-$E_F$ electronic structure of \ttise\ in the normal phase, i.e. in the absence of CDW fluctuations.
The constant energy map (left picture) consists of a hole pocket at $\Gamma$ ($\bar{\Gamma}$ in the surface BZ) and three symmetry equivalent electron pockets at $L$ and $L'$ ($\bar{M}$ and $\bar{M}'$ in the surface BZ) in planes parallel to the 
$\Gamma M K$ plane ($\bar{\Gamma}\bar{M}\bar{K}$ in the surface BZ). Schematically, the corresponding near-$E_F$ dispersions (right picture) involve three Se $4p$-derived valence bands having their maximum at $\Gamma$ and one Ti $3d$-derived conduction band having its minimum at $L$.

At RT already, strong fluctuations of the CDW phase are present \cite{CercellierPRL}, so that a gap opens between the valence and conduction bands, shifting the valence band below $E_F$.
Only the conduction bands remain in the neighbourhood of $E_F$. Therefore, to study the temperature dependence of the electronic structure of \ttise, we focus on the situation at $\bar{M}$. At LT, the most prominent feature attesting the CDW phase is seen at $\bar{M}$ in the form of the backfolded valence band. It is located well inside the occupied states, below the conduction band. Its maximum shifts to higher binding energies with decreasing temperature. Fig. \ref{PRB_Tscan_fig_2} shows photoemission intensity maps at the temperatures of 13K, 172K and 288K, corresponding to situations well below $T_c$, near $T_c$ and well above $T_c$, respectively.
Well below $T_c$, the backfolded valence band is intense and clearly distinct from the conduction band. 
In fact, two contributions can be resolved at high binding energies representing two backfolded valence bands. The conduction band provides some intensity just below $E_F$. 
Near $T_c$, the backfolded valence band is less intense and closer to the conduction band, which has gained intensity below $E_F$. Well above $T_c$, the backfolded valence band mixes up with the conduction band and only some residual intensity shows up below the conduction band, precursor of the CDW and the backfolded valence band.

\begin{figure}
\centering
\includegraphics[width=8.5cm]{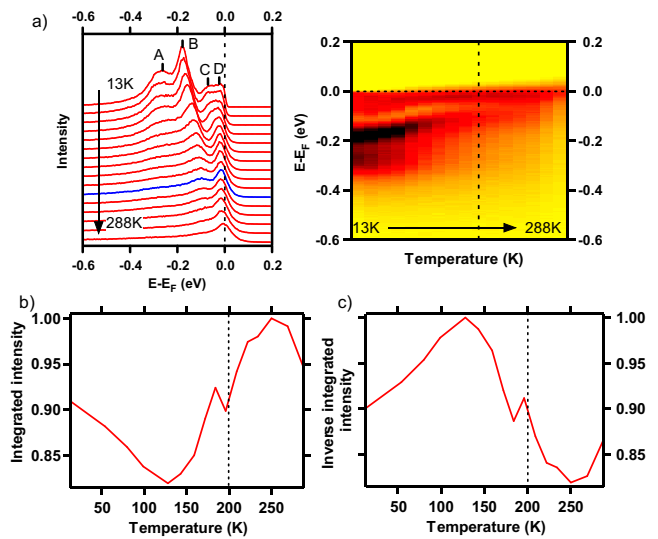}
\caption{\label{PRB_Tscan_fig_3} (color online)
(a) Left: EDCs measured exactly at $\bar{M}$, as a function of temperature (for the blue EDC, $T=200$K). Right: the corresponding false color plot, with the dashed line at $T=200$K. (b) Integrated intensity around $E_F$ (over $\pm50$ meV) for each measured photoemission intensity map, as a function of temperature. (c) Inverse integrated intensity.
}
\end{figure}

To understand the origin of the CDW phase of \ttise\ and its influence on the ARPES spectra, we have performed a detailed series of such measurements over a wide temperature range. From each intensity map, the central energy distribution curve (EDC) (situated at $\bar{M}$ exactly) is extracted, allowing to plot the waterfall of Fig. \ref{PRB_Tscan_fig_3}(a) (left). The blue EDC was measured at 200K, at $T_c$. The conduction band, labelled $C$, is clearly recognized just below $E_F$. At LT, a new peak, labelled $B$, develops below the conduction band and is identified as the valence band backfolded to $\bar{M}$. With decreasing temperature, it shifts to higher binding energies and becomes much more intense. This increase in spectral weight is mainly balanced by a decrease in spectral weight of the original valence band at $\bar{\Gamma}$ (not shown here, see reference \cite{MonneyPRBTheo} for a comprehensive discussion). Looking more carefully one sees that another peak, labelled $A$, also develops below this backfolded band, which can be identified with a second (spin orbit split) valence band backfolded from $\bar{\Gamma}$ to $\bar{M}$. 
Contribution $D$ is close to $E_F$ and is resolved here for the first time to our knowledge (similar measurements performed on other samples confirm this result). The splitting between contribution $C$ and $D$ is of the order of $\sim 50$ meV and as the temperature increases, contribution $D$ rapidly disappears. We identify this peak as the quasiparticle peak originating from the coupling of the conduction band to phonons \cite{HengsbergerElPh}. Contribution $C$ is then the incoherent peak which mostly follows the dispersion of the original conduction band. This phenomenon will be described in more details in a forthcoming paper \cite{MonneyElPh}.

Fig. \ref{PRB_Tscan_fig_3}(a) (right) shows a false color plot of this waterfall, emphasizing thus the temperature evolution of the position and intensity of the backfolded valence band. The vertical dotted line indicates $T_c$. The intensity of the photoemission intensity maps (as those shown in Fig. \ref{PRB_Tscan_fig_2}) is integrated in the vicinity of $E_F$ ($\pm 50$ meV around $E_F$) and plotted as a function of temperature in Fig. \ref{PRB_Tscan_fig_3}(b). This procedure provides us with an estimate of the temperature dependence of the electron density $n$ , which we can then relate to the particular behaviour of the resistivity of \ttise. 
Indeed, the electron density $n$ of the occupied states in the conduction band participating in the conductivity $\sigma$ according to the Drude formula, can be obtained by 
\begin{equation}
n\propto\int d^3k\int_{-\Delta\omega}^{+\Delta\omega} d\omega A_c(\vec{k},\omega)n_F(\omega).\nonumber
\end{equation}
Here $n_F(\omega)$ is the Fermi-Dirac distribution. $A_c(\vec{k},\omega)$ is the spectral function at $L$, which is measured by photoemission \cite{DamascelliPE}. Therefore, limiting the energy integration to the neighbourhood of the Fermi energy with $\Delta\omega=50$ meV permits to focus on the contribution of the conduction band. In Fig. \ref{PRB_Tscan_fig_3}(c), the inverse of the curve in Fig. \ref{PRB_Tscan_fig_3}(b) represents then an approximation of the corresponding resistivity (using the Drude formula not considering changes in relaxation time and mass). It displays qualitatively the behaviour of the measured resistivity curve \cite{DiSalvoSuperLatt,LevyResist}, with a sharp increase around $T_c$ and a decrease at lower temperatures (Fig. \ref{PRB_Tscan_fig_1}(a)). 

%

\begin{figure}
\centering
\includegraphics[width=8.5cm]{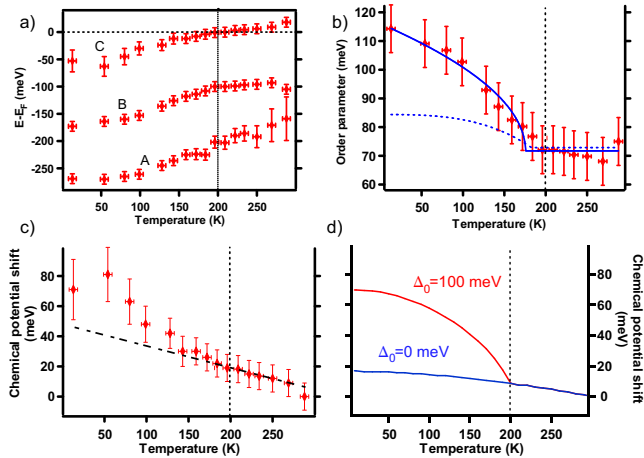}
\caption{\label{PRB_Tscan_fig_4}
(a) Position of contributions A,B,C as a function of temperature, obtained by fitting EDCs, and with their error bars. (b) Order parameter of the exciton condensate phase. The continuous (blue) line is a fit with a mean-field-like function. The dashed (blue) line is a similar fit, after subtraction of the chemical potential shift (see text for more details). (c) Chemical potential shift obtained from our data. The dotted-dashed line represents the extrapolation of a linear fit to the high temperature part. (d) Calculated chemical potential shift from previous work \cite{MonneyPhysicaB} (see text).
}
\end{figure}

To get more information from the data of Fig. \ref{PRB_Tscan_fig_3}(a), fitting EDCs as a function of temperature is necessary. 
Contribution $D$ complicates the fitting procedure of the EDCs at the lowest temperatures only, where it appears. Therefore, we adopt the following strategy. For EDCs at the lowest temperatures, we start by adjusting a Lorentzian to the contribution C alone (this approximation results in larger error bars) and subtract it from the EDC. Then, contributions A and B are fitted separately by two Lorentzians. 
At higher temperatures, the situation is simpler as fitting with three Lorentzians is possible at once for the whole EDC. Fitting is done in an iterative way, meaning that parameters of the previous fit (i.e. with a lower temperature) are used as an initial guess. The Fermi-Dirac cutoff is included in the fit too.
Fig. \ref{PRB_Tscan_fig_4}(a) displays the position of contributions $A$, $B$ and $C$ as a function of the temperature, namely the two backfolded valence bands and the conduction band respectively. Above $\sim 200$K, when the peaks start to merge together, the fitting procedure becomes less accurate. This is reflected by the larger error bars. All three bands undergo a shift towards $E_F$ as the temperature increases, with the largest change below $T_c$. Moreover, at RT, the center of the conduction band at $\bar{M}$ (close to $L$) is at 18 meV above $E_F$, meaning that the band lies in the unoccupied states at RT. The goal of these fits is to extract the temperature dependent order parameter from the experimental data within the exciton condensate phase model. The result is shown in Fig. \ref{PRB_Tscan_fig_4}(b) as explained in the next section.

\section{Discussion of the experimental data}

We now adopt the exciton condensate phase mechanism as the origin of the CDW transition to go further in our analysis and to understand our experimental results. In the exciton condensate phase, at LT, the CDW naturally appears from the condensation of excitons, which are bound pairs of holes from the valence band and electrons from the conduction bands. The non-zero center-of-mass momentum of the excitons, which is the distance between $\Gamma$ and $L$, gives rise to the CDW, as a purely electronic mechanism. This CDW phase is characterized by a non-zero order parameter $\Delta$, similar to that of the BCS theory (see reference \cite{MonneyPRBTheo} for a rigorous derivation). We have obtained the spectral function within this framework to calculate photoemission intensity maps. It gives us the renormalized dispersions in the CDW phase, as well as their spectral weight. We must emphasize that in our minimal model, we considered only the topmost valence band and the three symmetry equivalent conduction bands at $L$, coupled together by the Coulomb interaction (these are the bands represented by thick lines in Fig. \ref{PRB_Tscan_fig_1}(c)). 

\begin{figure}
\centering
\includegraphics[width=8.5cm]{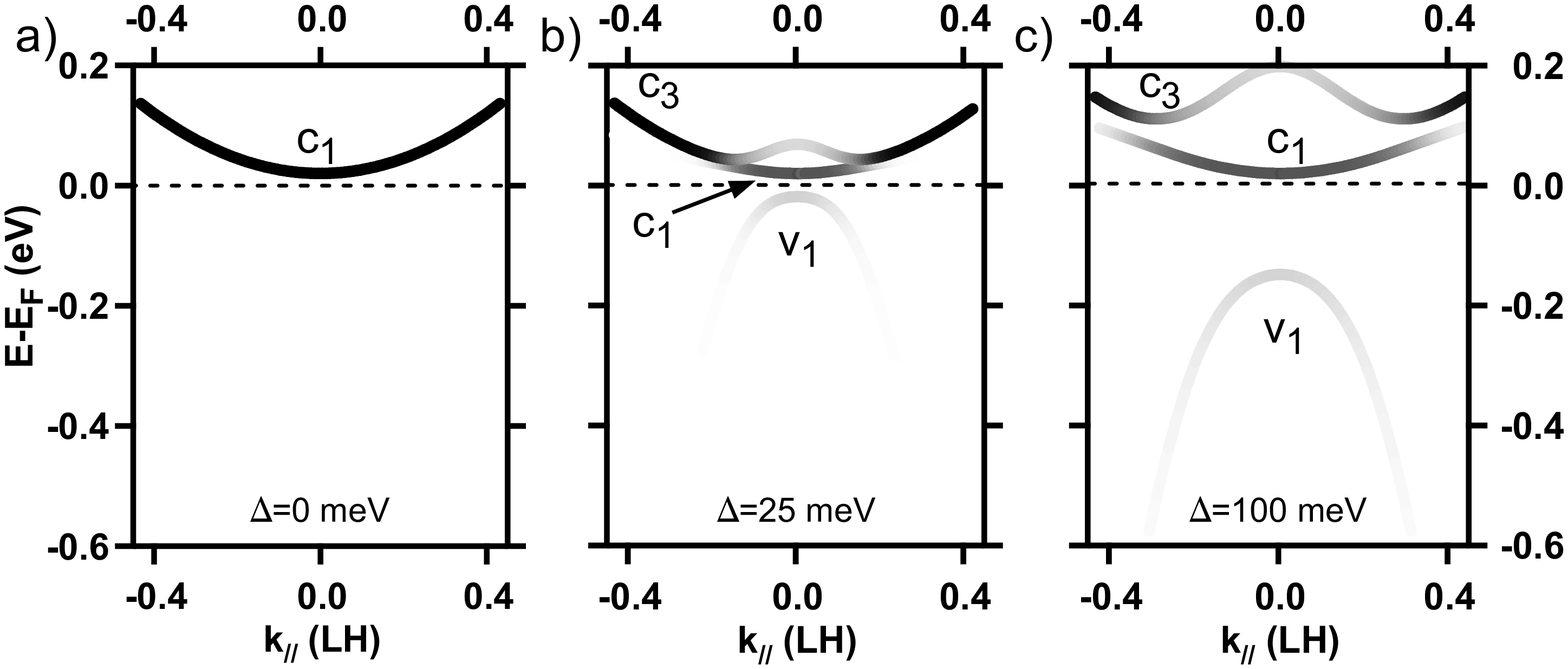}
\caption{\label{PRB_Tscan_fig_5} Renormalized dispersions at $L$ calculated by the spectral function derived for the exciton condensate phase for (a) the normal phase, $\Delta=0$ meV, (b) a weak CDW phase, $\Delta=25$ meV, and (c) a well-established CDW phase, $\Delta=100$ meV. The spectral weight is rendered by the grayscale (the darker the colour, the higher the spectral weight).
}
\end{figure}

Fig. \ref{PRB_Tscan_fig_5} shows calculated dispersions at $L$ and along the $LH$ direction, for different values of the order parameter $\Delta$ and with their spectral weight depicted with a grayscale. In the normal phase (graph (a)), $\Delta=0$ meV, only the original conduction band $c_1$ exists, slightly above $E_F$, with a unit spectral weight. For $\Delta=25$ meV (graph (b)), spectral weight is transferred from the original conduction band $c_1$ not only into the backfolded valence band $v_1$ which appears just below $E_F$, but also into a symmetry equivalent conduction band $c_3$ backfolded (from another $L$ point) above the original one. When $\Delta$ increases to 100 meV (graph (c)), the backfolded valence band $v_1$ is shifted further away from $E_F$, at higher binding energy. Above $E_F$, more spectral weight is transferred into the backfolded conduction band $c_3$.
As we can guess from this series of graphs, the order parameter $\Delta$ does not influence the minimum position of the conduction band exactly at $L$, provided the chemical potential is fixed. This was confirmed more precisely in previous work\cite{MonneyPRBTheo}. Therefore, in that model, any displacement of this band at $L$ will signify a chemical potential shift. 
Fig. \ref{PRB_Tscan_fig_4} (c) shows the chemical potential shift obtained with the shift of the position of the conduction band (peak C in Fig.  \ref{PRB_Tscan_fig_4} (a)) with respect to its RT value, as a function of temperature. The dashed-dotted line mimics a linear behaviour, revealing that this chemical potential shift is probably composed of two components. 

To get furter insight, we refer to a previous paper, where we addressed the question of the temperature dependence of the exciton condensate phase of \ttise\ by feeding the model with a temperature dependent  order parameter having a given mean-field form \cite{MonneyPhysicaB}. That study permitted us to visualize in the calculated spectra the temperature dependent spectroscopic signature of the exciton condensate phase and to understand the essential role played by the backfolded band to evaluate the order parameter. We also computed (not self-consistently) the chemical potential shift induced by the CDW transition, reproduced here in Fig. \ref{PRB_Tscan_fig_4} (d), as well as the change in the charge carrier density, which characterizes the insulating character of the transition. More precisely, it was shown that a semimetal like \ttise\ 
exhibits a chemical potential shift which is quasi-linear with temperature (blue curve in Fig. \ref{PRB_Tscan_fig_4} (d)), when no CDW transition takes place (if $\Delta=0$ meV over the whole temperature range). This is in fact a 
consequence of the Fermi-Dirac distribution on the two bands of the semimetal with non-equal band masses.
Once the excitonic transition sets in, the chemical potential shift deviates from this quasi-linear behaviour at $T_c$ and increases even more (red curve in Fig. \ref{PRB_Tscan_fig_4} (d)) due to the redistribution of spectral weight. The measured chemical potential in Fig. \ref{PRB_Tscan_fig_4}(c) displays a similar trend with good agreement. 

\begin{figure}
\centering
\includegraphics[width=9cm]{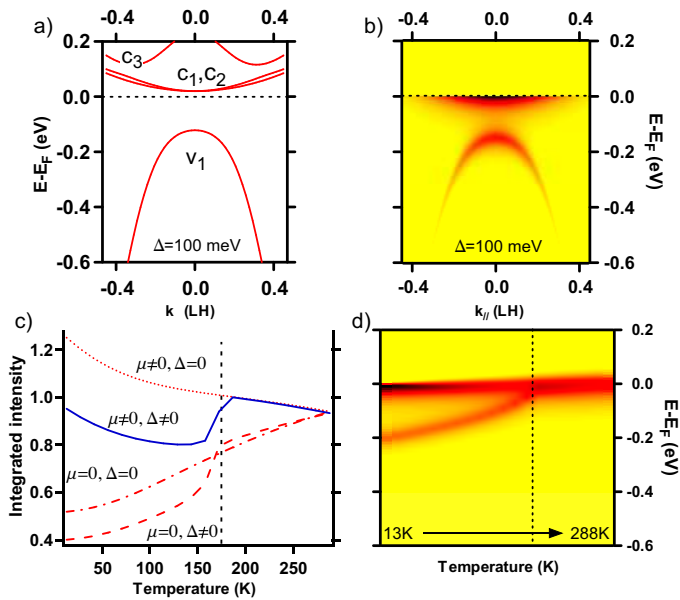}
\caption{\label{PRB_Tscan_fig_6}
(a) Near $E_F$ dispersions around $L$ of \ttise\ calculated  within the exciton condensate phase model for an order parameter $\Delta=100$ meV. (b) Corresponding (calculated) photoemission intensity map (as a false color plot), where the spectral weight has been added to the dispersions. (c) Integrated intensity around $E_F$ (over $\pm50$ meV) for each calculated photoemission intensity map, as a function of temperature. (d) False color plot made of the collection of calculated EDCs at different temperatures and at $L$ for the case of a non-zero chemical potential shift and a non-zero order parameter.
}
\end{figure}

In the exciton condensate model adapted to \ttise, once $\Delta\neq 0$ meV, new bands, that are the direct manisfestation of the CDW, develop at high symmetry points. To clearly show these different contributions, we plot in Fig. \ref{PRB_Tscan_fig_6}(a) the renormalized dispersions of Fig. \ref{PRB_Tscan_fig_5}(c) without the spectral weight information. We see that in fact four bands (the band labelled $c_2$ carries a negligible spectral weight and was therefore not visible in Fig. \ref{PRB_Tscan_fig_5}) appear near $L$ in the CDW phase. Their positions are complicated functions of $\Delta$. However, exactly at $L$, the situation simplifies drastically to
\begin{eqnarray}\label{eqn_positions}
E_{v_1}(\Delta)&=&E_{c_1}-\frac{E_G}{2}-\frac{1}{2}\sqrt{E_G^2+12\Delta^2}\nonumber,\\ E_{c_1}(\Delta)&=&E_{c_1},\qquad E_{c_2}(\Delta)=E_{c_1},\nonumber\\ E_{c_3}(\Delta)&=&E_{c_1}-\frac{E_G}{2}+\frac{1}{2}\sqrt{E_G^2+12\Delta^2} ,
\end{eqnarray}
with $E_{c_1}=18$ meV, the position of the conduction band at RT, and $E_G:=E_{c_1}-E_{v_1}=-12$ meV, the gap between the valence and conduction bands in the normal phase, which in our case is an overlap ($E_{v_1}=30$ meV is the position of the valence band for $\Delta=0$ meV taken from Ref. \cite{CercellierPRL}). These relationships are of great interest since they provide a simple way to extract the temperature dependence of the order parameter $\Delta(T)$ from the position of the backfolded bands. 

In our case, as mentioned above, only the backfolded valence band $E_{v_1}(\Delta)$ is useful, since the band $c_3$ cannot be observed by photoemission. 
Also, it must be emphasized that calculations leading to relation \ref{eqn_positions} have been done in a mean-field approximation, meaning that this relation is valid only below $T_c$. Nonetheless, we use it above $T_c$ to get an idea of the strength of fluctuations already identified in Fig. \ref{PRB_Tscan_fig_2} at 288K, where some diffuse intensity remains at about 100 meV below the conduction band, interpreted as a precursor of the emergence of the backfolded valence band and confirming this hypothesis. 

Inverting thus the first equation in (\ref{eqn_positions}) and inserting the temperature dependent position of the valence band of Fig. \ref{PRB_Tscan_fig_4}(a) ($E_{v_1}(\Delta)$, contribution $B$, which corresponds to $v_1$ in \ref{PRB_Tscan_fig_5}(c)) results in the data points of Fig. \ref{PRB_Tscan_fig_4}(b). For comparison, they are superimposed on a mean-field-like order parameter of the form
\begin{equation}\label{eqn_OP}
\Delta(T)=\Delta_0 \sqrt{1-\left(\frac{T}{\tilde{T}_c}\right)^\alpha} + \Delta_{{\rm off}}
\end{equation}
fitted to the experiment (continuous blue line), with $\alpha=1$. One sees immediately that the critical temperature extracted from the ARPES data $\tilde{T}_c=175$K is smaller than that determined from resistivity measurements\cite{DiSalvoSuperLatt} $T_c=200$K. For ensuring the best agreement with the data, an offset value for the order parameter $\Delta_{{\rm off}}=72$ meV has to be added above $T_c$, to model roughly the strong fluctuations of incoherent electron-hole pairs above $T_c$. Once the temperature decreases below $T_c$, the order parameter displays a clear increase in a mean-field fashion. This is a strong indication of the macroscopic condensation of coherent excitons. At the lowest temperature the order parameter reaches the value of $\Delta(T=0\text{K})\cong 120$ meV (extrapolated from $\Delta(T=13\text{K})=116$ meV). 

One may argue that part of the shift of the backfolded valence band, which is used to derive the order parameter curve, is due to the chemical shift of the electronic band structure. Indeed, both effects affect each other and a global treatment of these effects should be applied. In fact, the order parameter depends on the density of charge carriers (through screening of the potential), depending themselves on the chemical potential shift.  However such an approach implies a self-consistent numerical calculation of the gap equation and of the chemical potential. Due to the anisotropy and the multiplicity of the conduction bands (the three symmetry equivalent conduction bands at the $L$ points), it turns out to be more complicated than for a simple BCS case and thus goes beyond the scope of the present work. Nonetheless, we superimpose on Fig. \ref{PRB_Tscan_fig_4}(b) a fit (dashed blue line) to the measured order parameter obtained with the position of the backfolded valence band $E_{v_1}(\Delta)$ from which the chemical potential shift of Fig. \ref{PRB_Tscan_fig_4}(c) has been fully subtracted. It thus represents a lower limit for the order parameter.
\\

With the analysis of our photoemission data in the framework of the exciton condensate phase, we revealed here two effects at play in the \ttise\ system, as a function of temperature (i.e., the shift of the chemical potential and the onset of the order parameter). To better understand their influence on the properties of this system, we insert them into our model, first separately and then together. 

Fig. \ref{PRB_Tscan_fig_6}(b) depicts a photoemission intensity map generated by the spectral function calculated in this model \cite{MonneyPRBTheo}. An order parameter $\Delta=100$ meV has been used in this particular case (corresponding to a CDW phase with strong excitonic effects). It has to be reminded that only the topmost valence band is considered in our model, meaning that only one backfolded valence band will appear at $L$. 
For comparison with Fig. \ref{PRB_Tscan_fig_3}(b), the intensity of such calculated photoemission intensity maps in the vicinity of $E_F$ ($\pm 50$ meV around $E_F$) is integrated and plotted as a function of temperature in Fig. \ref{PRB_Tscan_fig_6}(c). The dashed dotted (red) curve corresponds to the case of a system with an order parameter $\Delta(T)=0$ meV and a chemical potential shift $\mu(T)=0$ meV. The integrated intensity around $E_F$ shifts to smaller values as a consequence of the narrowing of the Fermi-Dirac distribution only. If excitonic effects are included ($\Delta_0=100$ meV, $\Delta_{{\rm off}}=20$ meV, $\tilde{T}_c=175$K, see Eqn. (\ref{eqn_OP})), one obtains the dashed (red) curve. Here, the loss of spectral weight in the conduction band, transferred to the backfolded valence band due to the CDW formation, induces an additional decrease of the integrated intensity below $T_c$. On the contrary, when a linear (with temperature) chemical potential shift, reaching $25$ meV at LT, is taken into account (with $\Delta=0$ meV), the trend changes to the dotted (red) curve. Indeed, the chemical potential then shifts the conduction band into the occupied states, increasing thus the integrated intensity around $E_F$. In \ttise, both effects compete against each other, so that the resulting behaviour is non-monotonic. 

When combining these two effects, the integrated intensity (blue curve) shows the same behaviour than the measurements plotted in Fig. \ref{PRB_Tscan_fig_3}(b). It has to be noticed that the set of values ($\Delta_0=100$ meV, $\Delta_{{\rm off}}=20$ meV, $T_c=175$K) giving the best agreement to our measurements is different from what can be inferred from experiment ($\Delta_0=48$ meV, $\Delta_{{\rm off}}=72$ meV, $T_c=175$K, used to fit the data in Fig. \ref{PRB_Tscan_fig_4}(b)). We attribute this discrepancy to the fact that our model does not reproduce all the features of the experiment and to the fact that the chemical potential shift and the order parameter influence each other and cannot simply be described separately (as explained above).
In Fig. \ref{PRB_Tscan_fig_6}(d), the corresponding calculated EDCs (at $L$) are displayed in the same way as in Fig. \ref{PRB_Tscan_fig_3}(a) (right). It reproduces the main features of the experiment, namely the appearance of the backfolded valence band with a high intensity transferred from the conduction band, which shifts slowly into the occupied states as the temperature decreases. However, the second backfolded valence band is of course missing (not included in the model) and the intensity of the backfolded valence band is smaller than that of the original conduction band, in contrast to the experiment.

\section{Further discussions}

An unsolved debate concerning \ttise\ is the configuration of the bands near $E_F$. In the present work, the conduction band at $\bar{M}$ (but close to $L$) and at RT is observed to be above $E_F$, at $E_{c_1}=18\pm 10$ meV. This is in apparent contradiction with our previous work \cite{CercellierPRL} (but in agreement with earlier ARPES studies \cite{PilloTiSe2,KiddTiSe2,RossnagelTiSe2}), where it has been measured below $E_F$, at $E_{c_1}=-40\pm 5$ meV. 
A possible explanation lies in the uncontrolled excess of Ti in these samples \cite{DiSalvoSuperLatt}. In our recent work \cite{MonneyPRBTheo}, in an attempt to reconcile density functional theory calculations (which predict a semimetallic configuration with a large overlap between the valence and conduction bands \cite{FangLDA}) with the experiment, we argued that the electrons of the excess Ti atoms, which are probably situated in the van der Waals gap (close to the Se atoms) 
would fill the hole pocket of the (Se$4p$-derived) valence band, shifting it into the occupied states. However, this picture may be too simplistic and the doping electrons may fill the conduction band and push it slightly below $E_F$. Then, a conduction band situated at $E_{c_1}=-40$ meV is simply the consequence of some excess Ti in the sample.

Concerning the position of the valence band, the situation is also unclear. Indeed, we have seen that at RT already, strong excitonic fluctuations appear to be present, resulting in a situation similar to the CDW phase with $\Delta\neq0$ meV, reminiscent to fluctuations in the pseudo-gap region of high temperature superconductors. In other words, according to our model, the position of the valence band at RT does not correspond to that of the valence band in the normal phase (i.e. with $\Delta=0$ meV), since it is already shifted below the conduction band by excitonic fluctuations (see reference \cite{MonneyPRBTheo} for more details). From calculations performed within our model, only the top of the valence band is affected. Therefore, we fitted the branches of the valence band dispersion at RT with a parabola, as an extrapolation towards its \textit{normal phase} dispersion \cite{CercellierPRL}. A position of $E_v=30$ meV for the valence band was determined, in contradiction with other ARPES studies, where excitonic effects were not considered.

An important question is now to see to which extent the precise position of the valence and conduction bands may change the conclusions of the present study.
We used the temperature dependent position of the valence band backfolded to $\bar{M}$ to extract the temperature dependence of the order parameter of the exciton condensate phase. This operation was performed with the help of relationship (\ref{eqn_positions}), where the RT position of the conduction band $E_{c_1}$ and the gap $E_G=E_{c_1}-E_{v_1}$ are parameters. The RT position of the conduction band $E_{c_1}$ has been determined in this work. $E_G$ depends also on the RT position of the valence band $E_{v_1}$ which was determined in previous work (see above), yielding an overlap of $E_G=-12$ meV and thus a semimetallic configuration. To test the influence of $E_{v_1}$ on this result, we also considered a semiconducting configuration, with the conduction band at the same position ($E_{c_1}=18$ meV) but a valence band completely in the occupied states at $E_{v_1}=-30$ meV and calculated the corresponding order parameter. The resulting fit (also done with the equation (\ref{eqn_OP})) is very similar to the continuous (blue) curve of Fig. \ref{PRB_Tscan_fig_4}(b), but the curve is shifted to lower energy values by $19$ meV. It turns out in fact that the shift of the order parameter curve varies almost linearly with the size of the gap $E_G$ (and that its shape hardly changes).

A non-trivial issue is to know how to interpret the order parameter obtained from the measurements (Fig. \ref{PRB_Tscan_fig_4}(b)) which clearly consists of two regimes. Below $T_c$, it increases in a mean-field fashion, which we understand as a macroscopic condensation of excitons. This is expected from the mean-field theory we used to derive equation \ref{eqn_positions}. Mean-field theory predicts a zero value for the order parameter above $T_c$. However, from our measurements and our procedure for extracting this order parameter, we find that it displays a finite value above $T_c$. This comes from the fact that a contribution from the backfolded valence bands (which is not negligible as can be seen in the EDCs of Fig. \ref{PRB_Tscan_fig_3}(a)) was necessary to fit the EDCs above $T_c$. We then interpret this finite order parameter above $T_c$ as the signature of strong electron-hole pair fluctuations, in a manner similar to the phase fluctuations of the complex order parameter for high-temperature superconductors in the pseudo-gap phase \cite{EmeryHTSCfluct}. 
We emphasize  that the numerical value of the order parameter above $T_c$ must be considered carefully, since we did not analyze this part of the data with the appropriate theory. Nevertheless, our approach is confirmed by the high residual intensity below the conduction band at 288K (see Fig. \ref{PRB_Tscan_fig_2}).

We now discuss the transition from the fluctuating to the macroscopic condensation regime. Three different possible cases can be compatible with our data. (i) In Fig. \ref{PRB_Tscan_fig_4}(b), we fit the experimental data with a function given by equation (\ref{eqn_OP}) (continuous line). This function describes a mean-field condensation starting at $T_c=175$K, sitting on a constant background $\Delta_{{\rm off}}$. In this case, incoherent excitons giving rise to fluctuations of the order parameter are present from RT to the lowest temperature (meaning that strong fluctuations produce above $T_c$ a pseudo-CDW phase throughout the whole sample, hiding the normal phase to photoemission) and below $T_c$ coherent excitons generated by the macroscopic condensation add themselves up to the incoherent ones. (ii) As another way to interpret the data of Fig. \ref{PRB_Tscan_fig_4}(b), one can imagine that at $T_c$ the macroscopic condensation suddenly converts all the incoherent excitons present above $T_c$ into coherent ones so that only coherent excitons exist below $T_c$. (iii) Finally, as an alternative to the second case, the conversion of incoherent excitons into coherent ones could be progressive, so that the macroscopic condensation would start at a critical temperature higher than what seems obvious in Fig. \ref{PRB_Tscan_fig_4}(b). In that sense, the nearly constant background of fluctuating excitons would hide the starting macroscopic condensation and the real critical temperature $T_c^*$ would be larger than what observed for the two other cases.
Discriminating between these three scenarios is a difficult task, which requires a theoretical understanding of the fluctuation regime above $T_c$. This goes beyond the exciton condensate phase model we already investigated for \ttise\ \cite{MonneyPRBTheo}. Recently, Ihle {\it et al.} studied the excitonic insulator phase within the extended Falicov-Kimball model, in order to understand the metal insulator transition of TmSe$_{0.45}$Te$_{0.55}$ \cite{IhleEI}. They have drawn the corresponding phase diagram which strongly suggests a crossover from a BCS (weak coupling) to a Bose-Einstein condensate (BEC) (strong coupling) phase, which appears in the case of a semimetallic and a semiconducting configuration, respectively. On the BEC side, above the critical temperature $T_c$ of the exciton condensation, they predict the existence of preformed excitons (which do not exist on the BCS side above $T_c$). In the context of our work, this gives a possible explanation to the existence of fluctuating electron-hole pairs above $T_c$, provided that \ttise\ displays a semiconducting configuration which is in contradiction to our conclusion that the valence band maximum is at $E_{v_1}=+30$ meV. 

Finally, the strength of the coupling can be then estimated with the well-known BCS relationship $2\Delta(T=0\text{K})/k_BT_c=12.5$ to $15.5$. This value is four times that of usual BCS systems, suggesting that a strong coupling is at play in \ttise\ to build excitons. This would not be surprising, since the pairing interaction is a weakly screened Coulomb interaction for excitons (rather than an overscreened one for Cooper pairs). 

\section{Conclusion}

To summarize, we have performed angle-resolved photoemission measurements of \ttise\ between 13K and 288K. We focussed on the situation at $\bar{M}$ (near $L$), where the conduction band represents the main contribution to the bandstructure near $E_F$. From its position obtained as a function of temperature, an important temperature dependent chemical potential shift is revealed. At LT, the valence band is backfolded from $\bar{\Gamma}$ to $\bar{M}$ as a direct manifestation of a transition towards a charge density wave phase. In the framework of the exciton condensate phase, its position is directly linked to the order parameter describing this phase. From our measurements, we are able to extract this temperature dependent order parameter. It shows a clear increase below the critical temperature $T_c$ of the transition, demonstrating exciton condensation in a mean-field manner. Most remarkably, it points towards a non-zero value above $T_c$, which we interpret as the signature of strong electron-hole fluctuations. 
However, a theoretical study of the above-$T_c$ fluctuations, applied to the bandstructure of \ttise, is still lacking. Finally, we integrated the near-$E_F$ spectral weight around $\bar{M}$ as a function of temperature. The inverse of the resulting curve exhibits a striking similarity with the anomalous resistivity of \ttise. On the basis of our model we are able to reproduce this behaviour qualitatively, provided that a temperature dependent chemical potential shift is included in addition to excitonic effects.

\begin{acknowledgments}
We thank L. Forr\`o for valuable discussions.
We wish to acknowledge the support of our mechanical workshop and electronic engineering team. This project was supported by the Fonds National Suisse pour la Recherche Scientifique through Div. II and the Swiss 
National Center of Competence in Research MaNEP. 
\end{acknowledgments}


\begin{thebibliography}{99}

\bibitem{FlorianThesis}
F. Clerc {\it et al.}, J. Phys.: Condens. Matter, Special Issue, July(2007).

\bibitem{MotizukiBook}
{\it Structural Phase Transitions in Layered Transition Metal Compounds}, edited by K. Motizuki (Reidel, Dordrecht, 1986).

\bibitem{AebiJElectSpect}
P. Aebi {\it et al.}, J. Electr. Spectr. {\bf 117}, 433 (2001).

\bibitem{MorosanNature}
E. Morosan {\it et~al.}, Nature Physics {\bf 2},  544  (2006).

\bibitem{FangTaS2}
L. Fang {\it et al.}, Phys. Rev. B {\bf 72}, 014534 (2005).

\bibitem{DiSalvoSuperLatt}
F.~J. Di~Salvo {\it et~al.}, Phys. Rev. B {\bf 14},  4321 (1976).

\bibitem{LevyResist}
F. L\'evy {\it et~al.}, J. Phys. C {\bf 12},  473 (1979).
  
\bibitem{HoltPhonSoft}
M. Holt, P. Zschack, H. Hong, M.Y. Chou and T.C. Chiang, Phys. Rev. Lett. {\bf 86},  3799  (2001).
   
\bibitem{CercellierPRL}
H. Cercellier {\it et~al.}, Phys. Rev. Lett. {\bf 99}, 146403 (2007).

\bibitem{RaschAdsorption}
J.C.E. Rasch, T. Stemmler, B. Muller, L. Dudy and R. Manzke, Phys. Rev. Lett. {\bf 101}, 237602 (2008).

\bibitem{ZungerDFT}
A. Zunger and A.J. Freeman, Phys. Rev. B {\bf 17}, 1839 (1978).
 
\bibitem{HuguesBJT}
H.P. Hughes, J. Phys. C {\bf 10}, L319 (1977).
         
\bibitem{KeldyshEI}
L.V. Keldysh and Y.V. Kopaev, Fiz. Tverd. Tela (Leningrad) {\bf 6} 2791 (1965) 
[Sov. Phys. Solid State {\bf 6}, 2219 (1965)].

\bibitem{JeromeBasis}
D. J\'erome {\it et~al.}, Phys. Rev. {\bf 158},  462  (1967).

\bibitem{MonneyPRBTheo} 
C. Monney {\it et~al.}, Phys. Rev. B {\bf 79}, 045116 (2009).

\bibitem{PilloTiSe2}
T. Pillo {\it et~al.}, Phys. Rev. B {\bf 61},  16213  (2000).

\bibitem{KiddTiSe2}
T.~E. Kidd, T. Miller, M.Y. Chou and T.C. Chiang, Phys. Rev. Lett. {\bf 88}, 226402  (2002).
   
\bibitem{RossnagelTiSe2}
K. Rossnagel, L. Kipp and M. Skibowski, Phys. Rev. B {\bf 65}, 235101 (2002).

\bibitem{DamascelliPE}
A. Damascelli, Physica Scripta {\bf T109}, 61 (2004).

\bibitem{HengsbergerElPh}
M. Hengsberger, D. Purdie, P. Segovia, M. Garnier and Y. Baer, Phys. Rev. Lett. {\bf 83}, 592 (1999).

\bibitem{MonneyElPh}
C. Monney {\it et al.}, unpublished.

\bibitem{MonneyPhysicaB} 
C. Monney {\it et~al.}, Physica B {\bf 404}, 3172 (2009).

\bibitem{FangLDA}
C.M. Fang, R.A. deGroot and C. Haas, Phys. Rev. B {\bf 56}, 4455 (1997).

\bibitem{EmeryHTSCfluct} 
V.J. Emery and S.A. Kivelson, Nature {\bf 374}, 434 (1995).  

\bibitem{IhleEI}
D. Ihle, M. Pfafferott, E Burovski, F.X. Bronold and H. Fehske, Phys. Rev. B {\bf 78},  193103  (2008).


\end{thebibliography}

\end{document}